\documentclass[11pt]{iopart}

\usepackage{chapterbib}
\usepackage{color}
\usepackage{graphics}      
\usepackage{graphicx}      
\usepackage{longtable}     
\usepackage{bm}            
\usepackage[colorlinks=true]{hyperref}
\usepackage{subfigure}
\usepackage{iopams}

\begin{document}

\title[1D Magnetization transport] {Experimental characterization of coherent
magnetization transport in a one-dimensional spin system}

\author {Chandrasekhar Ramanathan$^{1}$, Paola Cappellaro$^2$, \\
Lorenza Viola$^1$, David G. Cory$^{3,4}$}
\ead{sekhar.ramanathan@dartmouth.edu} 
\address{$^1$Department of Physics and Astronomy, Dartmouth
College, Hanover, NH 03755, USA}

\address{$^2$Department of Nuclear Science and Engineering,
Massachusetts Institute of Technology, Cambridge, MA 02139, USA}

\address{$^3$Department of Chemistry and Institute for Quantum
Computing, University of Waterloo, Waterloo ON N2L 3G1, Canada}
  
\address{$^4$Perimeter Institute for Theoretical Physics, Waterloo, ON
N2L 2Y5, Canada}

\date{\today}

\begin{abstract}
We experimentally characterize the non-equilibrium, room-temperature
magnetization dynamics of a spin chain evolving under an effective
double-quantum Hamiltonian.  We show that the Liouville space
operators corresponding to the magnetization and the two-spin
correlations evolve 90 degrees out of phase with each other, and drive
the transport dynamics.  For a nearest-neighbor-coupled $N$-spin
chain, the dynamics are found to be restricted to a Liouville operator
space whose dimension scales only as $N^2$, leading to a slow growth
of multi-spin correlations. Even though long-range couplings are present in the real
system, we find excellent agreement between the analytical predictions
and our experimental results,
confirming that leakage out of the restricted Liouville space is slow
on the timescales investigated.
Our results indicate that the group velocity of the magnetization is
$6.04\pm0.38$ $\mu$m/s, corresponding to a coherent transport over
$N\approx 26$ spins on the experimental timescale.  As the
double-quantum Hamiltonian is related to the standard one-dimensional
XX Hamiltonian by a similarity transform, our results can be directly
extended to XX quantum spin chains, which have been extensively
studied in the context of both quantum magnetism and quantum
information processing.
\end{abstract}

\maketitle


\section{Introduction}

Solid-state spin systems provide an attractive test-bed to study both
equilibrium and non-equilibrium quantum many-body dynamics, and have
recently emerged as a promising platform for quantum simulation
\cite{Roumpos-2007,Buluta-2009}.  One-dimensional (1D) spin systems,
in particular, are of special interest as they connect to an important
class of problems in condensed-matter physics \cite{Giamarchi-2009},
and have been suggested as quantum wires to coherently transfer
quantum information across distant nodes in a quantum computer and
distributed quantum architectures \cite{Bose-2003,Friesen-2007}.
Transport properties, and magnetization transport in particular, have
been extensively investigated theoretically by both the
condensed-matter community and more recently in the context of
spintronics and emerging nano-device applications \cite{Meier-2003}.

The class of 1D spin-$1/2$ XY Hamiltonians \cite{Lieb-1961,Korepin-1997,Mattis-2006}, which are
exactly solvable via a Jordan-Wigner mapping onto a system of
non-interacting spinless fermions, play an archetypal role in condensed-matter physics.  
In this case, a local magnetic
disturbance is known to propagate down the chain scatter-free with a
constant velocity ({\em ballistic transport}), rather than diffusively
spreading from the site of the disturbance and eventually decaying
({\em diffusive transport}).  From a quantum communication
perspective, the mapping to free fermions has proved crucial to also
allow a quantum state to be transported down the chain --- which thus
acts as a quantum conduit or channel (see for example \cite{Kay-2010}
for a recent review).  While at zero temperature, within linear
response theory, integrable quantum models are typically associated
with ballistic transport, a coexistence with and/or crossover to
diffusive behavior may be possible more generally, for instance in the
presence of non-local conserved quantities \cite{Sirker-2009} or of
couplings to an environment
\cite{Esposito-2005,Znidaric-2010,Eisler-2011}.  Despite
significant progress, a satisfactory understanding of the conditions
leading to ballistic versus diffusive transport is as yet lacking, with a
number of unresolved questions remaining, in particular, in relation
to the impact of finite and infinite temperatures \cite{Shastry-2008},
the role of non-integrability, and its interplay with frustration
\cite{Meisner-2004,Santos-2009}.  As a result, experimentally
characterizing the transport properties and physical mechanisms in
low-dimensional spin systems remains important from both fundamental
and applied standpoints.

The long coherence times afforded by nuclear spins and the ability to
access a large Hilbert space, in conjunction with the superb level of
control available over spin degrees of freedom, make solid-state
nuclear magnetic resonance (NMR) an excellent setting for exploring
the coherent dynamics of a (nearly) isolated quantum many-body system
\cite{Roumpos-2007,Cho-2005,Alvarez-2010,Rigol-2010} as well as the
statistical physics of equilibrating spin systems
\cite{Bruschweiler-1997,Waugh-1998,Morgan-2008}.  In this work, we
employ solid-state NMR techniques to characterize the room-temperature
magnetization dynamics of a (quasi) 1D spin system during coherent
evolution under an effective Double Quantum (DQ) Hamiltonian, which is
directly related to the isotropic XY (XX henceforth) Hamiltonian by a
similarity transformation.  In particular,
we find that the experimental results are in very good agreement with
those predicted by the free-fermionic solutions, indicating that
integrability-breaking perturbations (due to longer range couplings)
have negligible effect on the timescale of the experiments, and enabling us to calculate the
transport velocity of the magnetization.  This is in marked contrast
to the diffusive behavior observed under the dipolar Hamiltonian in
previous experiments on 3D spin networks
\cite{Zhang-1998a,Boutis-2004}, where spin diffusion arises from
unitary dynamics under the high-field (secular) dipolar Hamiltonian.
While experimentally the high degree of isolation from the surrounding
environment during evolution was demonstrated by the observation of
``polarization echoes'' upon reversing the sign of the dipolar
Hamiltonian \cite{Zhang-1992}, quantum chaoticity was explicitly
invoked theoretically as a mechanism for diffusion
\cite{Levstein-1998,Pastawski-2000}.


Beside providing additional insight into the mechanisms underlying
coherent transport in {\em isolated 1D many-body quantum systems}, in
the context of quantum information transport it is important to stress
that the nuclear spin chains we study here are initially in a {\em
highly mixed} quantum state.  The importance of relaxing
initialization constraints is being increasingly appreciated within
the quantum communication community (see in particular
\cite{Cappellaro-2011} and references therein).  We thus expect this
study to also be of direct relevance to a number of other quantum
platforms where mixed-state spin chains are naturally encountered,
such as phosphorus defects in silicon nanowires~\cite{Ruess07},
quantum dots~\cite{Wang04, Nikolopoulos04}, molecular
semiconductors~\cite{Petit90} and solid-state defects in diamond or
silicon carbide~\cite{Weber10,Wrachtrup10}.


\section{Experimental Methods and Results}

The $^{19}$F spins in a crystal of fluorapatite (FAp --
Ca$_5$(PO$_4$)$_3$F) have long been used to experimentally approximate
a nearest-neighbor coupled 1D spin system (see for example
\cite{Engelsberg73,Cho93,Cho96,Oliva08,Zhang-2009}).  In a 3D lattice
of dipolar-coupled nuclear spins, every pair of spins is coupled with
an interaction strength (between spin $j$ and spin $\ell$) $d_{j\ell} =
(\mu_0/16\pi) (\gamma^2 \hbar /r_{j\ell}^3) (1-3\cos^2\theta_{j\ell})$
\cite{Slichter}, where $\gamma$ is the gyromagnetic ratio of fluorine,
$r_{j\ell}$ is the distance between nucleus $j$ and $\ell$, and
$\theta_{j\ell}$ is the angle between $\vec r_{j\ell}$ and the
$z$-axis (along which the external magnetic field is applied).  The
geometry of the spin system is reflected in the distribution of the
$d_{j\ell}$ couplings.  In FAp, the $^{19}$F nuclei form linear chains
along the $c$-axis, each one surrounded by six other chains. The
distance between two intra-chain $^{19}$F nuclei is $r=3.442$~\AA \;
whereas the distance between two cross-chain $^{19}$F nuclei is
$R=9.367$~\AA. The largest ratio between the strongest intra- and
cross- chain couplings ($\approx 40$) is obtained when the crystalline
$c$-axis is oriented parallel to the external field. In this
orientation, the fluorine spins may be treated as a collection of many
identical 1D chains with only nearest-neighbor (NN) couplings, with a
coupling strength $d \equiv (\mu_0/8\pi) \gamma^2/r^3 = 8.17 \times
10^3$ rad/s.

In this NN approximation, the high-field secular dipolar Hamiltonian
of a single chain is given by \mbox{$\mathcal{H}_\mathrm{Dip} =
\frac{1}{2}\sum_{i=1}^{N-1} d \left(3\sigma_{i}^z\sigma_{i+1}^z -
\vec{\sigma}_{i} \cdot \vec{\sigma}_{i+1}\right)$}, where
$\sigma^{\alpha}$ ($\alpha=x,y,z$) are the Pauli operators.  Starting
from $\mathcal{H}_\mathrm{Dip}$ and using suitable multiple-pulse
sequences \cite{Yen-1983,Ramanathan-2003}, we can experimentally
implement an effective DQ Hamiltonian, given by
\vspace*{-0.09in}
\begin{equation}
\mathcal{H}_{\mathrm{DQ}} = \frac{1}{2}\sum_{i}^{N-1} d
\left(\sigma_{i}^x\sigma_{i+1}^x - \sigma_{i}^y\sigma_{i+1}^y\right).
\vspace*{-0.08in}
\label{eq:HDQ}
\end{equation} 
Formally, the DQ Hamiltonian is related to the standard XX Hamiltonian
by the similarity transformation $U_{\mathrm{DQ}}^{\mathrm{XX}} =
\exp(-i\pi/2 \sum_i' \sigma_x^i)$, where the sum is restricted to
either even or odd spins.  While $I_z \propto \sum_i I_z^i$ is
conserved under the XX Hamiltonian, the corresponding conserved
quantity under the DQ Hamiltonian is $\tilde{I}_z \propto
U_{\mathrm{DQ}}^{\mathrm{XX}} I_z U_{\mathrm{DQ}}^{\mathrm{XX}\dag}$.
This results in the inversion of sign of the local magnetic
disturbance at every alternate site as it moves down the chain, as
will be seen later (see Figure~\ref{fig:DQinterferences}).


In order to implement the DQ Hamiltonian, we used a standard 8-pulse
sequence applied on-resonance to the $^{19}$F Larmor frequency
\cite{Yen-1983}.  This $8$-pulse sequence, ${\tt S}={\tt C}\cdot
\bar{\tt C}\cdot \bar{\tt C}\cdot {\tt C}$, may be understood in terms
of a simpler $2$-pulse cycle ({\tt C}, and its time-reversed version
$\bar{\tt C}$), which also simulates the DQ Hamiltonian.  The
primitive pulse cycle is given by ${\tt C}=[\frac{\Delta}{2} X \Delta'
X \frac{\Delta}{2}]$, where $\Delta'= 2\Delta + w$, $\Delta$ is the
delay between pulses and $w$ is the width the $\pi/2$ pulse (denoted
by $X$), applied about the $x$-axis.  The dynamics in the presence of
the pulse sequence can be expressed in terms of a time-independent
effective Hamiltonian $\overline{\mathcal{H}}_{DQ}$,
\begin{eqnarray}
    U^{x}_{MQ}(t)& = &\mathcal{T}\exp\left(-i\int_0^t [\mathcal{H}_{dip}
    + {\cal H}^{x}_{rf}(s)]ds\right) = e^{-
    i\overline{\mathcal{H}}_{DQ}t},
\label{MQpropagator}
\end{eqnarray}
where $\mathcal{T}$ denotes time-ordering operator, $\hbar=1$, and
${\cal H}_{rf}^{x}(t)$ is the time-dependent Hamiltonian describing
the rf-pulses along the $x$-axis.  This sequence implements the
Hamiltonian of Eq. (\ref{eq:HDQ}) to the lowest order in an average
Hamiltonian theory sense \cite{Haeberlen76}.

In an inductively detected NMR experiment where we measure quadrature
signals from the $x$ and $y$ components of the precessing
magnetization, the observed signal is $S(t)=\zeta \langle
\sigma^{-}(t) \rangle =\zeta
\mathrm{Tr}\left\{\sigma^{-}\rho(t)\right\}$, where $\sigma^{-} =
\sum_j \sigma_{j}^- = \sum_j (\sigma_{j}^x - i \sigma_j^y)/2$, and
$\zeta$ is a proportionality constant. The only terms in $\rho(t)$
that yield a non-zero trace, and therefore contribute to $S(t)$, are
angular momentum operators such as $\sigma_{j}^+$, which are
single-spin, so-called ``single-quantum coherences'' in the language
of multiple-quantum (MQ) NMR \cite{Emsley-1993}.  In a standard MQ experiment that is
used to characterize the many-spin dynamics of a nuclear spin system,
the uncorrelated thermal initial states of the spins are allowed to
evolve under a Hamiltonian such as the DQ Hamiltonian that generates
the multi-spin dynamics.  The multi-spin character of the states
is indirectly encoded by a collective rotation of the spins and the DQ
evolution is then reversed to convert the many-spin states back to
single spin terms that can be detected.

\begin{figure}[thb]
\centering
\includegraphics[scale=0.5]{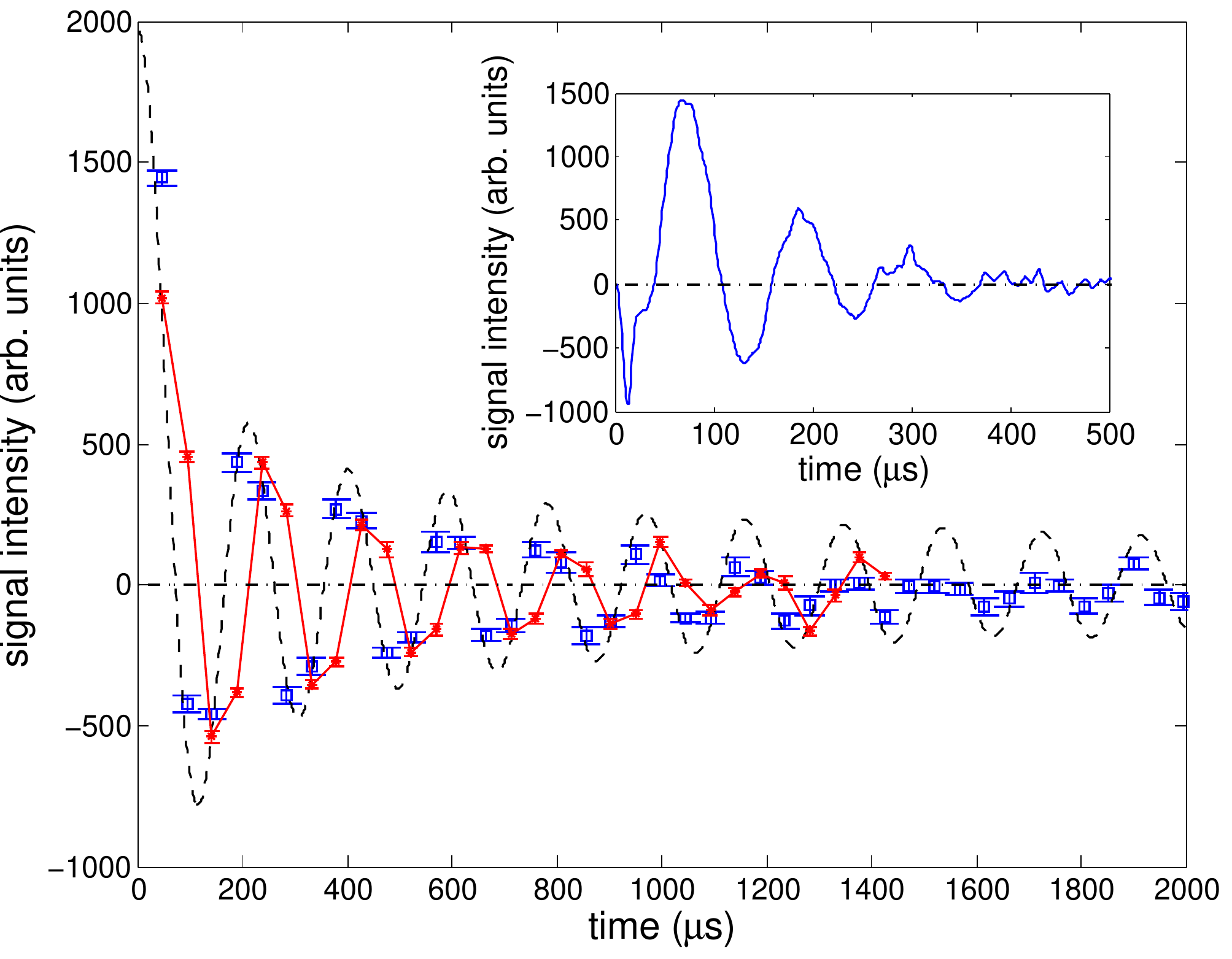}
\caption{Evolution of the thermal initial state under
$\mathcal{H}_{\mathrm{DQ}}$.  The blue squares correspond to the
collective magnetization ($\sum_i \sigma_z^i$) and the red asterisks
to the two-spin correlations ($\sum_{i} (\sigma_x^i\sigma_y^{i+1} +
\sigma_y^i\sigma_x^{i+1})$).  The dashed black line is the best fit of
the observed magnetization to the analytical model described
later. The solid red line is a guide to the eye. The inset shows the
NMR FID of FAp.}
\label{fig:fidcomp}
\end{figure}

Our experiments were performed at room temperature in a 7T vertical
bore NMR magnet on a single FAp crystal with its
$c$-axis aligned to the external field.  The measured T$_1$ of the
fluorine spins was 300 s (many orders of magnitude larger than the
timescales explored experimentally here).  The length of the $\pi/2$
pulse used was $w=1.06$ $\mu$s and $\Delta$=2.9 $\mu$s.  The {\em
thermal (equilibrium) state} of the spins in high field and high
temperature is highly mixed, and is given by $\rho_{\mathrm{th}}
\approx \mathbf{1}- \epsilon \sum_i\sigma_z^i$, where $\epsilon \sim
10^{-5} \ll 1$.  The identity is unchanged under unitary
transformations, and does not contribute to the signal $S(t)$.  Thus,
it is only the {\em deviation} of the density operator from the
identity that gives rise to an observable signal.  The constant
$\epsilon$ then becomes a scaling parameter and its value does not
affect the details of the experiment (as long as the high temperature
approximation remains valid).

Figure~\ref{fig:fidcomp} shows the observed evolution of the
collective magnetization ($\propto \sum_i \sigma_z^i$ -- blue squares)
and the two-spin correlations ($\propto \sum_i (\sigma_x^i
\sigma_y^{i+1} + \sigma_y^i \sigma_x^{i+1})$ -- red asterisks) under
$\mathcal{H}_{\mathrm{DQ}}$, starting from $\rho_{\mathrm{th}}$.  The
error bars were estimated from the standard deviation of a signal-free
region of the time-domain data.  The evolution time was incremented by
increasing the number of cycles from 1 to 40 (30 in the two-spin
correlation readout).

In contrast to typical NMR experiments which involve the DQ
Hamiltonian \cite{Yen-1983}, no evolution reversal was performed
before signal detection in our experiments.  In order to minimize
receiver dead-time effects, a solid echo was used to read out the
(single-spin) magnetization terms.  The two-spin terms were read out
using a $\pi/4$ pulse to generate a dipolar echo (with an appropriate
phase cycle).  The combination of a $\pi/4$ pulse followed by
evolution under the dipolar Hamiltonian refocuses a portion of the
two-spin correlations back to single-spin coherences that can then be
detected \cite{Slichter}.  The importance of the single-spin and
two-spin operators in driving the transport dynamics can be seen from
the following equations (the short time dynamics are further discussed 
in Appendix B):
\begin{equation}
\hspace*{-0.5in} \frac{d}{dt} \sigma_1^z= \frac{id}{2}
\left(\sigma^x_1\sigma^y_2 + \sigma^y_1\sigma^x_2\right), 
\hspace*{0.5cm} \frac{d}{dt}
\left(\sigma_1^x\sigma_2^y+\sigma_1^y\sigma_2^x\right)= -d
\,\sigma_2^z + \mathrm{3\:spin\:terms}.
\end{equation}
As a result, the single-spin and two-spin terms are observed to evolve
90 degrees out of phase with each other during DQ evolution.  The
inset shows the observed NMR Free Induction Decay (FID), which
corresponds to the evolution of $\sum_i \sigma_x^i$ under
$\mathcal{H}_{\mathrm{Dip}}$.  While the FID decays in about 350
$\mu$s, due to the creation of multi-spin correlations
\cite{Cho-2005}, the magnetization oscillations persist for up to
$\approx 1.5$ ms under $\mathcal{H}_{\mathrm{DQ}}$, indicating that
high-order spin correlations develop quite slowly
\cite{Ramanathan-2005}.

\begin{figure}[t]
\centering
\includegraphics[scale=0.5]{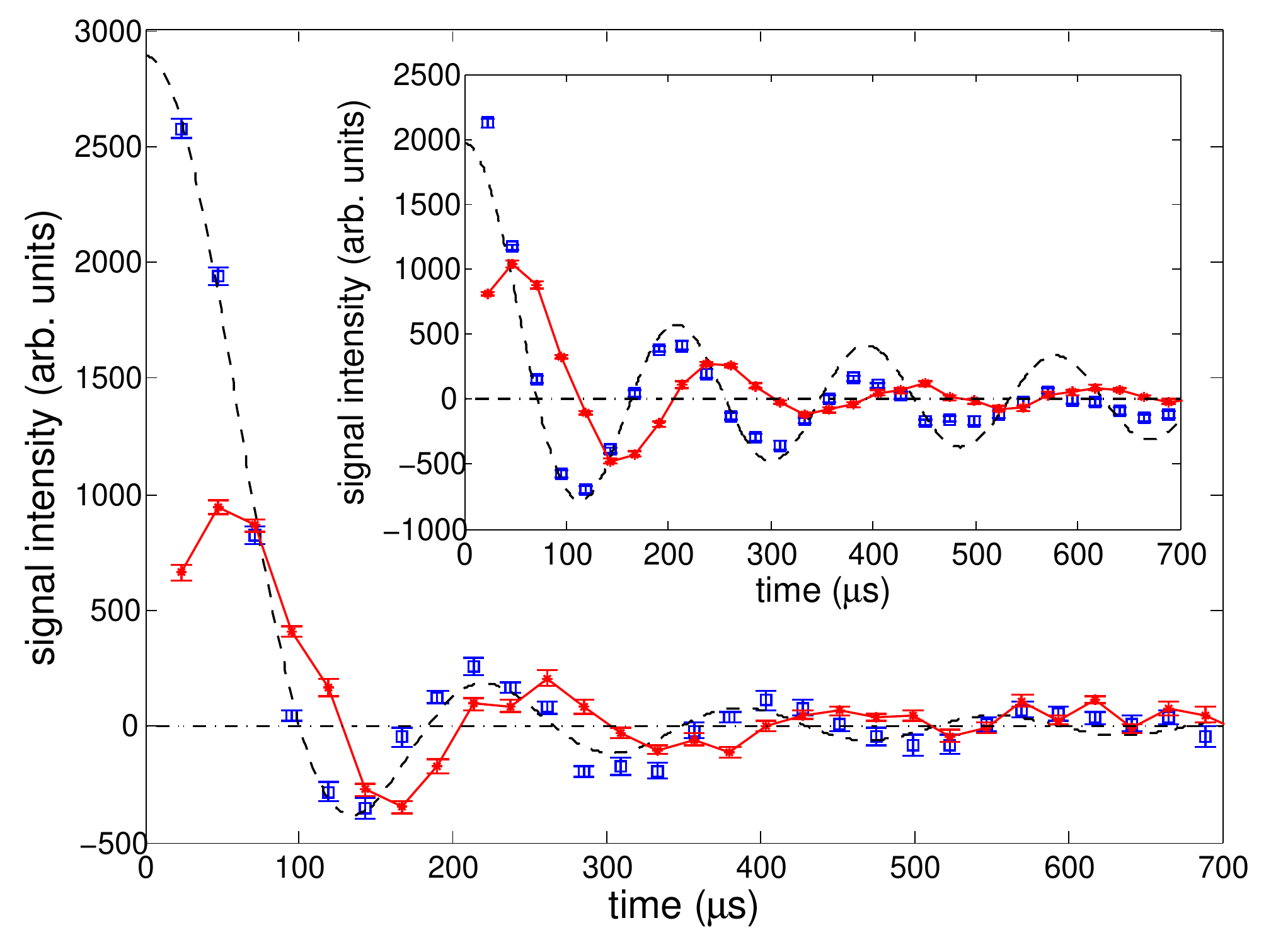}
\caption{Evolution of the end-polarized initial state under
$\mathcal{H}_{\mathrm{DQ}}$.  The blue squares correspond to $\sum_i
\sigma_z^i$, while the red asterisks correspond to $\sum_{i}
(\sigma_x^i\sigma_y^{i+1} + \sigma_y^i\sigma_x^{i+1})$.  Inset shows
the 4-pulse experiment on the thermal initial state.  The dashed black
lines are the best fits of the magnetization dynamics to the
analytical model.  }
\label{fig:4spin}
\end{figure}

It is also possible to experimentally prepare initial states of the
form $\rho_{1N} \approx \mathbf{1} - \epsilon(\sigma_1^z +
\sigma_N^z)$ \cite{Cappellaro-2007a}, in which the polarization is
localized at the ends of the chain.  We call this the
\emph{end-polarized state}.  Figure~\ref{fig:4spin} shows the DQ
evolution observed for $\rho_{1N}$.  Here, we implemented
$\mathcal{H}_{\mathrm{DQ}}$ using the first 4 pulses of the above
sequence (${\tt C}\cdot \bar{\tt C}$) to achieve better temporal
sampling of the signal, though this introduces first-order errors in
the resulting average Hamiltonian.  The alternating sign of the local
magnetization during spin transport of the end-polarized state results
in a rapid attenuation of the amplitudes of the observed signal (the
8-pulse version of the experiment is included in Appendix A).  For
comparison, the inset shows the 4-pulse version of the thermal state
experiment.

\section{Theoretical Analysis and Interpretation}
\subsection{Fermionic Model}

Both the NN XX \cite{Lieb-1961} and NN DQ
\cite{Cappellaro-2007a,Feldman-1997} Hamiltonians are well known to be
analytically solvable in 1D, by means of a Jordan-Wigner mapping onto
a system of free fermions.  We invoke this mapping to interpret the
experimental results.  Starting from an end-polarized initial state
$\rho_j = \mathbf{1} - \sigma_j^z$, we once again take note of the
fact that the identity does not evolve, and focus our attention on the
deviation term $\sigma_j^z$. The deviation density operator at time
$t$ under the DQ Hamiltonian is given by
\cite{Cappellaro-2007a,Cappellaro-2007b}:
\begin{eqnarray}
\vspace*{-0.05in} \rho_j^{\mathrm{DQ}}(t) & \hspace*{-1mm}=
\hspace*{-1mm}& \frac{1}{N+1}\sum_{k,h=1}^N \sin(kj)\sin(hj) \times \\
& & \hspace*{-0.6in} \left[\left(a_k^\dag a_h + a_h^\dag a_k \right)
\cos\psi_{kh}(t) -i\left(a_k^\dag a_h^\dag - a_h a_k\right)
\sin\psi_{kh}(t) \right], \nonumber
\vspace*{-0.05in}
\label{eq:end-dq1}
\end{eqnarray}
where $ a_k = \sqrt{\frac{2}{N+1}}\sum_{h=1}^{N} \sin(kh) c_h$, $c_h =
-\prod_{l=1}^{h-1} (\sigma_z^l)\sigma_{-}^h$ are canonical fermionic
operators, and $\psi_{kh}(t) = 2dt[\cos(k) +\cos(h)]$, and we have
assumed open boundary conditions on the chain.  In order to
characterize the evolution of the individual spin operators, we can
express the evolved state $\rho_j^{\mathrm{DQ}}(t)$
in terms of the $c_h$ operators.
This yields:
\begin{eqnarray}
\vspace*{-0.05in}
\rho_j^{\mathrm{DQ}}(t) &\hspace*{-1mm}=\hspace*{-1mm}& (-1)^{j-1} \Big\{
\!\!\!\sum_{p-q \; \in \; \mathrm{even}} \hspace*{-2mm} 
\left(c_p^\dag c_q + c_q^\dag c_p \right)A_{j,q}(t)A_{j,p}(t) \nonumber 
\\ & \hspace*{-1mm}+ \hspace*{-1mm}&  
\!\!\!\!\! 
\sum_{p-q \; \in \; \mathrm{odd}} \hspace*{-2mm}
i^{p-q} \left(c_p^\dag c_q^\dag - c_q c_p \right)A_{j,q}(t)A_{j,p}(t)\Big\},
\vspace*{-0.05in}
\label{eq:end-dq3}
\end{eqnarray}
where the time-dependent amplitudes $A_{j,q}$ read
\begin{eqnarray}
\vspace*{-0.05in}
A_{j,q}(t) & = & \sum_{m=0}^\infty i^{2m\tilde{N}} \left[
i^{\delta}J_{2m\tilde{N}+\delta}(2dt) -
i^{\Sigma}J_{2m\tilde{N}+\Sigma}(2dt) \right] \nonumber \\ & &
\hspace*{-0.5in} +\sum_{m=1}^\infty i^{2m\tilde{N}} \left[
i^{-\delta}J_{2m\tilde{N}-\delta}(2dt) -
i^{-\Sigma}J_{2m\tilde{N}-\Sigma}(2dt) \right]
\vspace*{-0.05in}
\label{eq:end-dq4}
\end{eqnarray}
with $\tilde{N} = N+1$, $\delta = q-j$, $\Sigma = q+j$, and
$J_n(\cdot)$ being the $n$-th order Bessel function of the first kind.
Similarly, for a thermal initial state $\rho_{\mathrm{th}} =
\mathbf{1} - \sum_j \sigma_z^j$, we find the deviation density operator at time $t$
\begin{eqnarray}
\vspace*{-0.05in} \rho^{\mathrm{DQ}}_{\mathrm{th}}(t) & =&
-\hspace*{-3mm}\sum_{p-q \; \in \; \mathrm{even}}\left(c_p^\dag c_q +
c_q^\dag c_p \right)A_{p,q}(2t) \nonumber \\ & & 
+ \sum_{p-q \; \in \;
  \mathrm{odd}} \left(c_p^\dag c_q^\dag - c_q c_p
\right)A_{p,q}(2t)\:.
\vspace*{-0.05in}
\label{eq:all-dq2}
\end{eqnarray}

Taking the even/odd constraint into account, the density operators in
Eqs.\ (\ref{eq:end-dq3}) and (\ref{eq:all-dq2}) are seen to belong to
a $N(N+1)/2$-dimensional operator subspace which {\em defines} the
Liouville space within which the transport occurs.  Mapping back to
spins, and assuming $ p \geq q$, we get:
\begin{equation}
\begin{array}{ll}
c_p^\dag c_q^\dag = \sigma_q^+\sigma_{q+1}^z\cdots\sigma_{p-1}^z
\sigma_p^+ ,  & \hspace*{0.1in} c_q c_p =
\sigma_q^-\sigma_{q+1}^z\cdots\sigma_{p-1}^z \sigma_p^-,  
\\ c_q^\dag c_p = \sigma_q^+\sigma_{q+1}^z\cdots\sigma_{p-1}^z \sigma_p^- , 
& \hspace*{0.1in} c_q^\dag c_q = \frac{1}{2}\left(\mathbf{1}-\sigma_q^z\right) .
\end{array}
\label{eq:end-dq7}
\end{equation}
The above quadratic scaling is the same that Fel'dman and coworkers
theoretically established for the XX Hamiltonian
\cite{Feldman-1998}\footnote{It is also interesting to observe that
the operators in Eq. (\ref{eq:end-dq7}) are the same as the string
operators in the spin-spin correlation functions
$S_{\mathrm{XX}}(q,p)$ defined in T.~S.~Cubitt and J.~I.~Cirac,
{Phys. Rev. Lett.} {\bf 100}, 180406 (2008).}.  While
$\rho_{\mathrm{th}}$ is a constant of the motion under the XX
Hamiltonian, evolution of the end-polarized state can be readily
expressed as
\begin{equation}
\vspace*{-0.05in}
\rho_j^{\mathrm{XX}}(t) = - \sum_{p,q} i^{p+q} \left(c_p^\dag c_q +
c_q^\dag c_p \right)A_{j,q}(t)A_{j,p}(t)  .
\vspace*{-0.05in}
\label{eq:end-xx}
\end{equation}


\subsection{Magnetization Dynamics and Transport Velocity}

We can re-examine the experimental data in Figs.\ 1 and 2 using the
analytical results outlined above.  The magnitude of the collective
spin magnetization is given by $ S(t) = \sum_{p=1}^N A_{1,p}^2(t)$ for
an end-polarized initial state, and by $S(t) = \sum_{p=1}^N
A_{p,p}(2t)$ for a thermal initial state.  The dashed black lines in
the figures are the best fits to these expressions, where we have
assumed that $N$ is sufficiently large that no boundary effects are
observed.  Thus we only used the $m=0$ term in Eq.\ (\ref{eq:end-dq4})
to calculate $A_{1,p}$ and $A_{p,p}$ to calculate $S(t)$ as shown
above.  For the end-polarized state we have also assumed that on the
timescale of the experiment the magnetization propagating from the two
ends have not had a chance to overlap, and we can thus describe them
as two independent chains.  This allows us to ignore the $A_{N,p}$
term.

Three fitting parameters were used: a scalar multiplier, the frequency
argument of the Bessel function, and an additive baseline constant.
The baseline constant was subtracted from the data shown in the two
figures, so that the oscillations are observed around zero.  For the
thermal initial state, the observed signal is seen to damp out at a
faster rate than expected from the model.  This is likely due to the
presence of longer-range couplings that have been ignored in the NN
model, as these long-range couplings lead to a leakage out of the
restricted Liouville space.  The obtained fitting frequencies yield $d
= 8.32 \times 10^3$ rad/s (8-pulse, thermal state), $d=8.52 \times
10^3$ rad/s (4-pulse, thermal state), $d = 8.71 \times 10^3$ rad/s
(8-pulse, end polarized state -- see Appendix A), and $d=9.56 \times
10^3$ rad/s (4-pulse, end polarized state), yielding an estimate $d =
8.78 \pm 0.55 \times 10^3$ rad/s.  The estimate is biased by the value
obtained in the 4-pulse experiment on the end-polarized state, where
the fitting frequency is seen to be too high at longer times (Fig.\
\ref{fig:4spin}).  However, the estimate is still in good agreement
with the values $d=8.3 \times 10^3$ rad/s, obtained from observing MQ
coherences \cite{Zhang-2009}, and $d=8.17 \times 10^3$ rad/s, obtained
from the known structure of FAp \cite{Cho96}.

In the thermodynamic limit, the magnitude of the magnetization
transported from site $j=1$ at time $t=0$ to site $j=n$ at time $t$ is given by
$P^\infty_{1,n}(t)=A^\infty_{1,n}(t)^2$, where
\begin{equation}
A^\infty_{1,n}(t)=i^{n-1}\frac{n
J_n(2dt)}{2dt}=i^{n-1}\frac{J_{n-1}(2dt)+J_{n+1}(2dt)}{2} \: \: .
\end{equation}
Using properties of the Bessel functions \cite{Abramowitz-1972}, it is
possible to show that
$\partial_t{A^\infty_{1,n}}=-d(A^\infty_{1,n+1}-A^\infty_{1,n-1})$.
If we define a continuous spatial variable $z=aj$, with $a$ equal to
the distance between two spins and $j$ being the spin number in the
chain, we can replace the finite difference with a spatial derivative
$\partial_t{A^\infty_{1,n}}=-2a d\partial_z{A^\infty_{1,n}}$.  Taking
the second derivative with respect to time we thus obtain a wave
equation for the transport amplitude:
\begin{equation}
\partial^2_t{A^\infty_{1,n}}= (2ad)^2\partial_z^2{A^\infty_{1,n}}.
\end{equation}
$P^\infty_{1,n}(t)$ also follows the same wave equation with velocity
$2ad$.  We can also calculate the group velocity of the spin system
directly from the dispersion relation for the DQ Hamiltonian, $\omega
(k)= 2d|\cos(k a)|$, which results in 
$v_g = \partial \omega/\partial k|_{k=\pi/2a} =2a d$.  
Using $d = 8.78 \pm 0.55 \times 10^3$ rad/s, and $a = 3.442$ \AA, we
obtain $v_g \approx 6.04 \pm 0.38$ $\mu$m/s.  This corresponds to a
displacement of $\approx 9.07 \pm 0.57$ nm in 1.5 ms or transport
across $N=26\pm 2$ spins.
The time
taken to travel a distance of $n$ lattice sites is $t = n/2d$.  At
this time $A_{1,n}^\infty = i^{n-1} J_n(n)$, and $P_{1,n}^\infty =
(-1)^{n-1} J_n^2(n)$.  For large $n$, $J_n(n)\sim \frac
1{\Gamma(2/3)}\left(\frac2{9n}\right)^{1/3}$ \cite{Abramowitz-1972}
and the magnitude of the polarization transported from spin 1 to $n$
scales as
\begin{equation}
P^\infty_{1,n}(n) \sim  n^{-2/3}.
\end{equation}
\noindent


\subsection{Evolution of Multi-spin Correlations}

\begin{figure}[htb]
\centering
\includegraphics[scale=0.5]{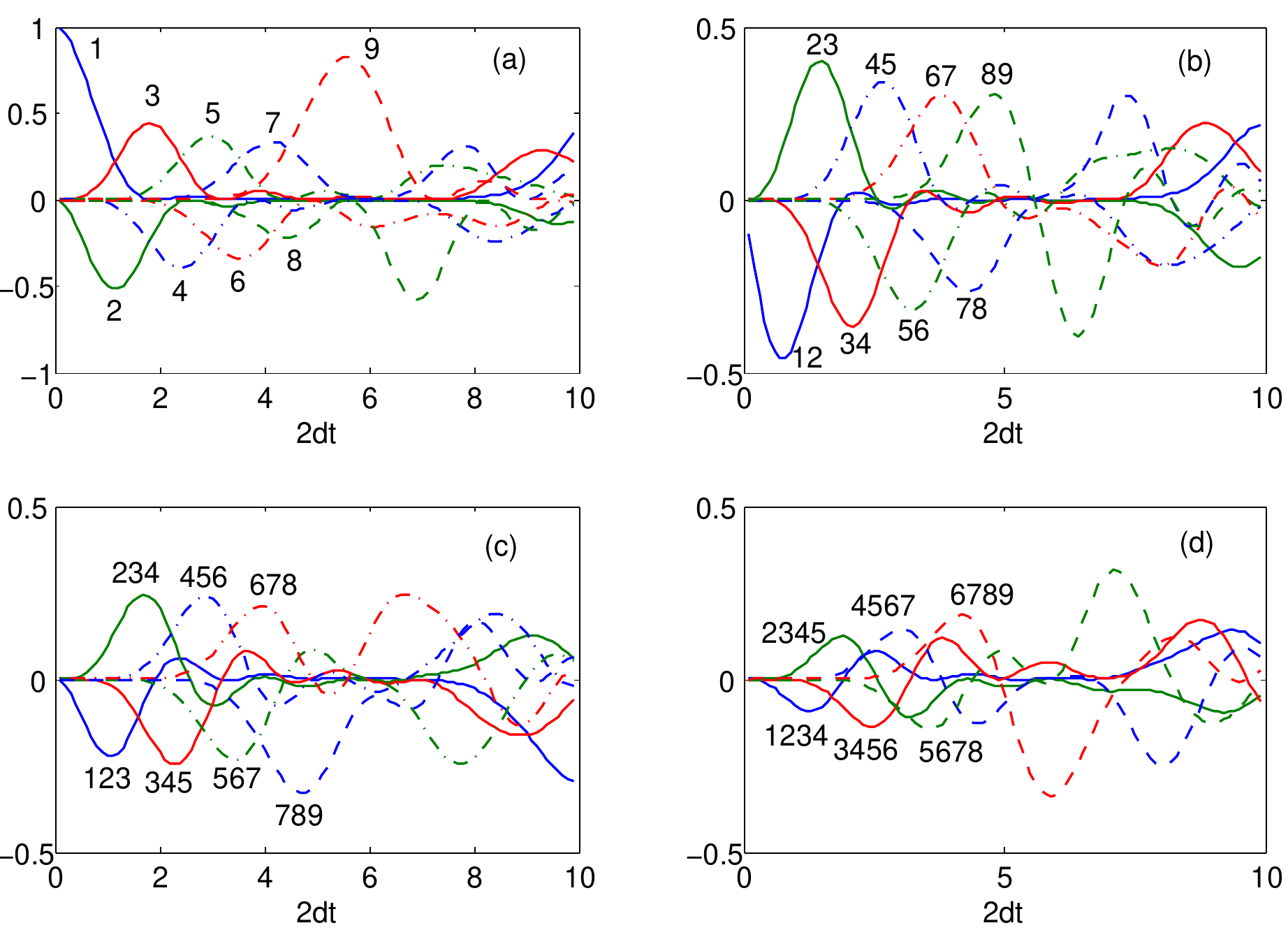}
\caption{This figure illustrates the evolution of the 1,2,3, and
4-spin correlations for a $N=9$ spin system initialized in the state
$\rho_1 = \mathbf{1} - \sigma_1^z$.  The amplitudes of different
product operators are obtained from Eqs.\ (3) and (4) in the main
paper.  (a) Single spin polarization terms $\sigma_n^z$.  The
amplitudes are given by $A_{1,n}^2(t)$. (b) Two spin correlation terms
$h^{(2)}_{n,n+1}$.  These are double quantum terms whose amplitudes
are given by $A_{1,n}(t) A_{1,n+1}(t)$. (c) Three spin correlation
terms $h^{(3)}_{n,n+1,n+2}$.  These are zero quantum terms whose
amplitudes are given by $A_{1,n}(t) A_{1,n+2}(t)$. (d) Four spin
correlation terms $h^{(4)}_{n,n+1,n+2,n+3}$.  These are double quantum
terms whose amplitudes are given by $A_{1,n} (t) A_{1,n+3}(t)$. Higher
order correlations are not shown.}
\label{fig:transp}
\end{figure}

Figure~\ref{fig:transp} shows the dynamics of the 1,2,3, and 4-spin
correlations for a $N=9$ spin chain initialized in the state $\rho_1=
\mathbf{1} - \sigma_z^1$.  These curves were generated using Eqs.\ (3)
and (4).  We can express $A_{1,q}$ as
\begin{eqnarray*}
\hspace*{-1in}A_{1,q} &= &\sum_{m=0}^{\infty}
i^{2m\tilde{N}+q-1}\left(\frac{2m\tilde{N}+q}{dt}\right)J_{2m\tilde{N}+q}(2dt)+
\sum_{m=1}^{\infty}
i^{2m\tilde{N}-q+1}\left(\frac{2m\tilde{N}-q}{dt}\right)J_{2m\tilde{N}-q}(2dt)
\end{eqnarray*}
where we have used the Bessel function identity $J_{\nu-1}(z) +
J_{\nu+1}(z) = \frac{2\nu}{z}J_\nu(z)$ \cite{Abramowitz-1972}.  The
sequential growth outlined above can easily be visualized in these
simulations.  The figures show that the presence of the boundary (at
$2dt_m \approx 5$, where $t_m\sim N/2d$ identifies the so-called
``mirror time'', see also next section) has a large effect on the spin
dynamics.  In Appendix B we show that additional insight into the
dynamics is obtained by calculating the short-time evolution, and
examining the growth of the multi-spin operators.

It is possible to experimentally characterize the growth of these
multi-spin correlations.  In a MQ experiment \cite{Yen-1983}, we
record the coherence orders by performing a collective rotation of the
state about the $z$-axis, and observing the resulting phase shifts.
The first row of Eq.~(\ref{eq:end-dq7}) shows the double-quantum
coherences created during NN DQ evolution, while the second row shows
the zero-quantum coherences and the polarization states.

\begin{figure}[htb]
\centering
\includegraphics[scale=0.5]{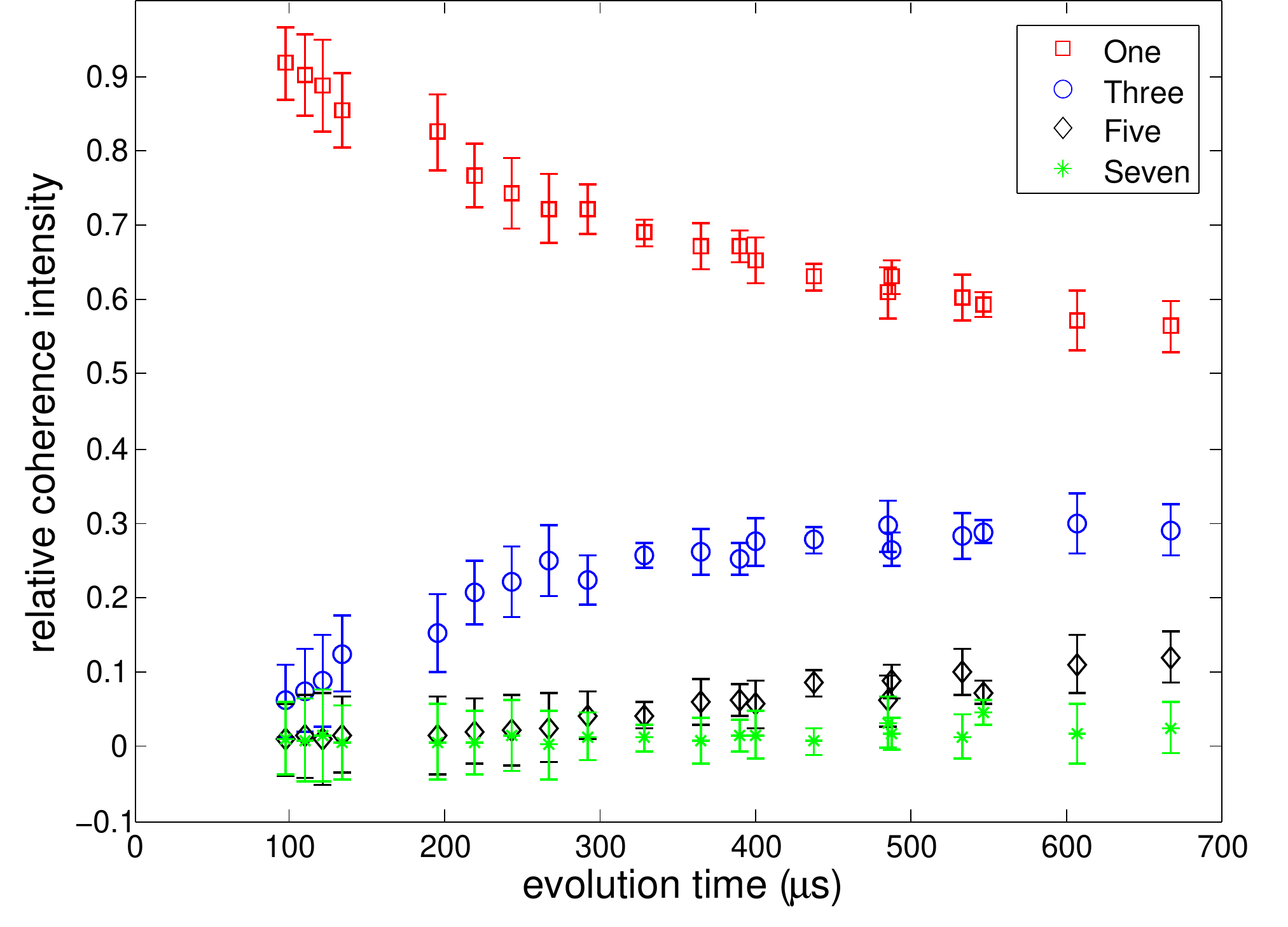}
\caption{Growth of $x$-basis coherences during DQ
evolution, starting from an end-polarized initial state.  Only
odd-order coherences are observed due to selection rules.}
\label{fig:endpolxb}
\end{figure}

While we cannot directly observe the high-spin correlations in Eq.\
(\ref{eq:end-dq7}), we can use so-called $x$-basis encoding techniques
to characterize the distribution of $p-q$ \cite{Ramanathan-2003},
thereby indirectly probing the growth of these terms.  For example,
the first term in Eq. (\ref{eq:end-dq7}) in the $x$-basis reads:
\begin{eqnarray*}
\hspace*{-1in}c_p^\dag c_q^\dag & = & \left[\sigma_q^{x}+
\frac{i}{2}(\sigma_q^{x+}+\sigma_q^{x-})\right] 
\frac{(\sigma_{q+1}^{x+} - \sigma_{q+1}^{x-})
\cdots (\sigma_{p-1}^{x+}-\sigma_{p-1}^{x-})}
{(2i)^{p-q-2}}\left[\sigma_p^{x}+\frac{i}{2}(\sigma_p^{x+}+\sigma_p^{x-})\right]\:,
\end{eqnarray*}
where $\sigma^{x\pm}_q = \sigma^y_q \pm i\sigma^z_q$
\cite{Zhang-1995}.  Here, a collective rotation of the system about
the $x$-axis results in overlapping binomial distributions of phase
factors whose highest order is $p-q$.  Higher-order coherences in the
$x$-basis are thus a signature of the presence of multi-spin
correlations.  Figure~\ref{fig:endpolxb} shows the relative $x$-basis
coherence intensities measured as a function of the DQ evolution time,
starting from the end-polarized state. The experiments were performed
using a 16-pulse implementation of the DQ Hamiltonian
\cite{Ramanathan-2003}, and both the number of cycles and the delay
$\Delta$ were varied.  It can be seen that following an initial rapid
creation of 3-spin correlations (and concomitant reduction in the
single spin term), the coherence orders change quite slowly.

\subsection{Mirror Times}

\begin{figure}[t]
\centering \includegraphics[scale=0.4]{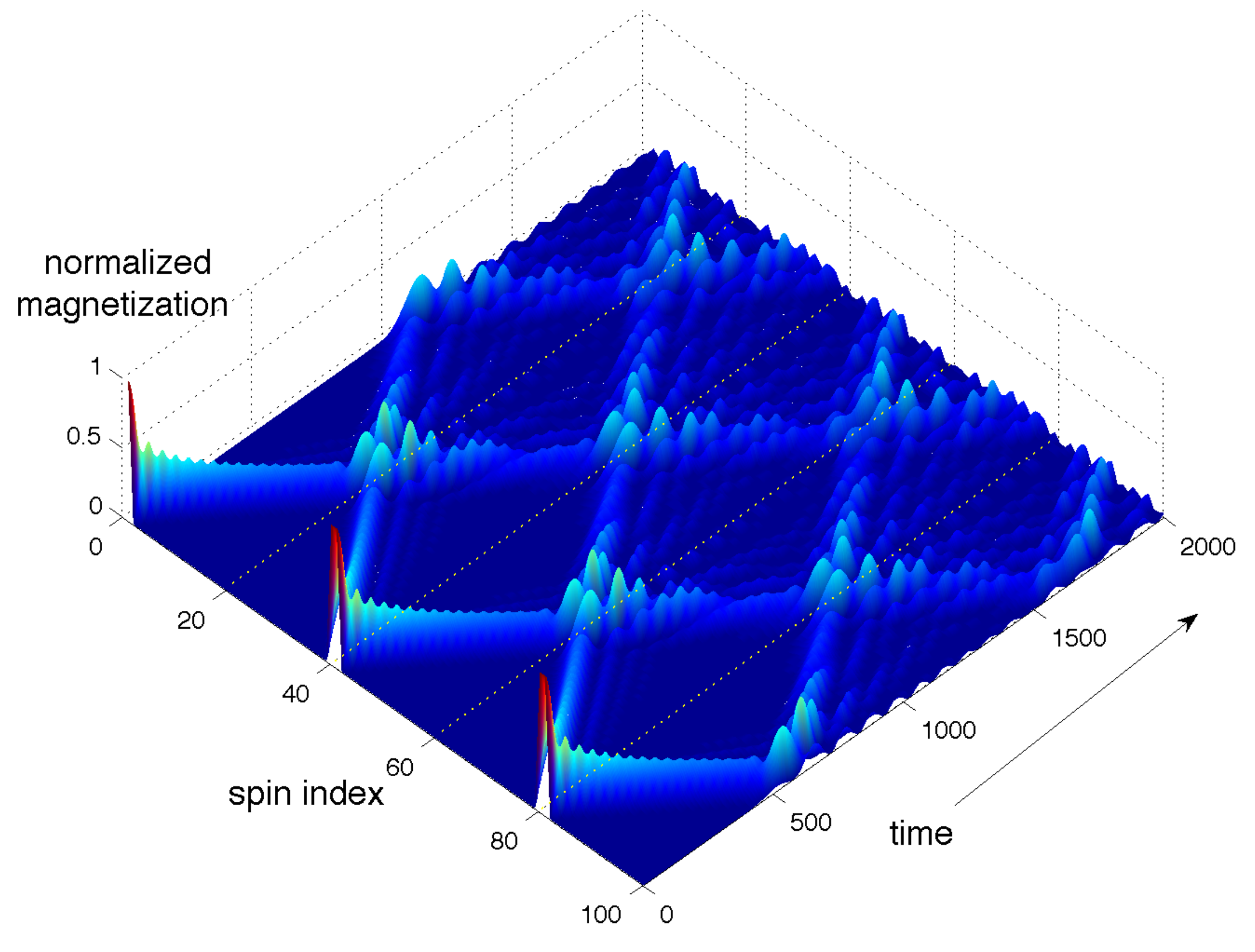}
\caption{Five copies of a 20-spin chain illustrate the origin of the
interference patters observed following reflection off the boundary
during evolution under the NN XX Hamiltonian. The time axis is in
units of $1/d$.}
\label{fig:XXinterferences}
\end{figure}

Equations~(3)--(7) also help us understand the source of the signal
enhancements observed near the boundaries of a finite spin chain.  The
sum over $m$ in Eq.\ (\ref{eq:end-dq4}) resembles a sum over an
infinite number of copies of the $N$-spin chain (consistent with the
periodic boundary conditions imposed by the sine transform
\cite{Cappellaro-2007b}).  At short times, only the lower-order Bessel
functions contribute, and each ``replica'' of the chain is independent
of the others.  At longer times, the adjacent copies begin to
interfere with each other as illustrated in Fig.
\ref{fig:XXinterferences}.  It is the interference between these terms
that is responsible for the {\em mirror times} observed in our
previous work \cite{Zhang-2009}, consistent with the interpretation in
terms of bouncing spin-wave packets and ``erratic'' dynamics put
forward for the XX model \cite{Feldman-1998}.  In particular, the
leading mirror term yields a factor of 4 increase in the magnitude of
the transferred magnetization.  A similar transport behavior has been
shown for pure states \cite{Banchi-2010}.

\begin{figure}[htb]
	\centering \includegraphics[scale=0.4]{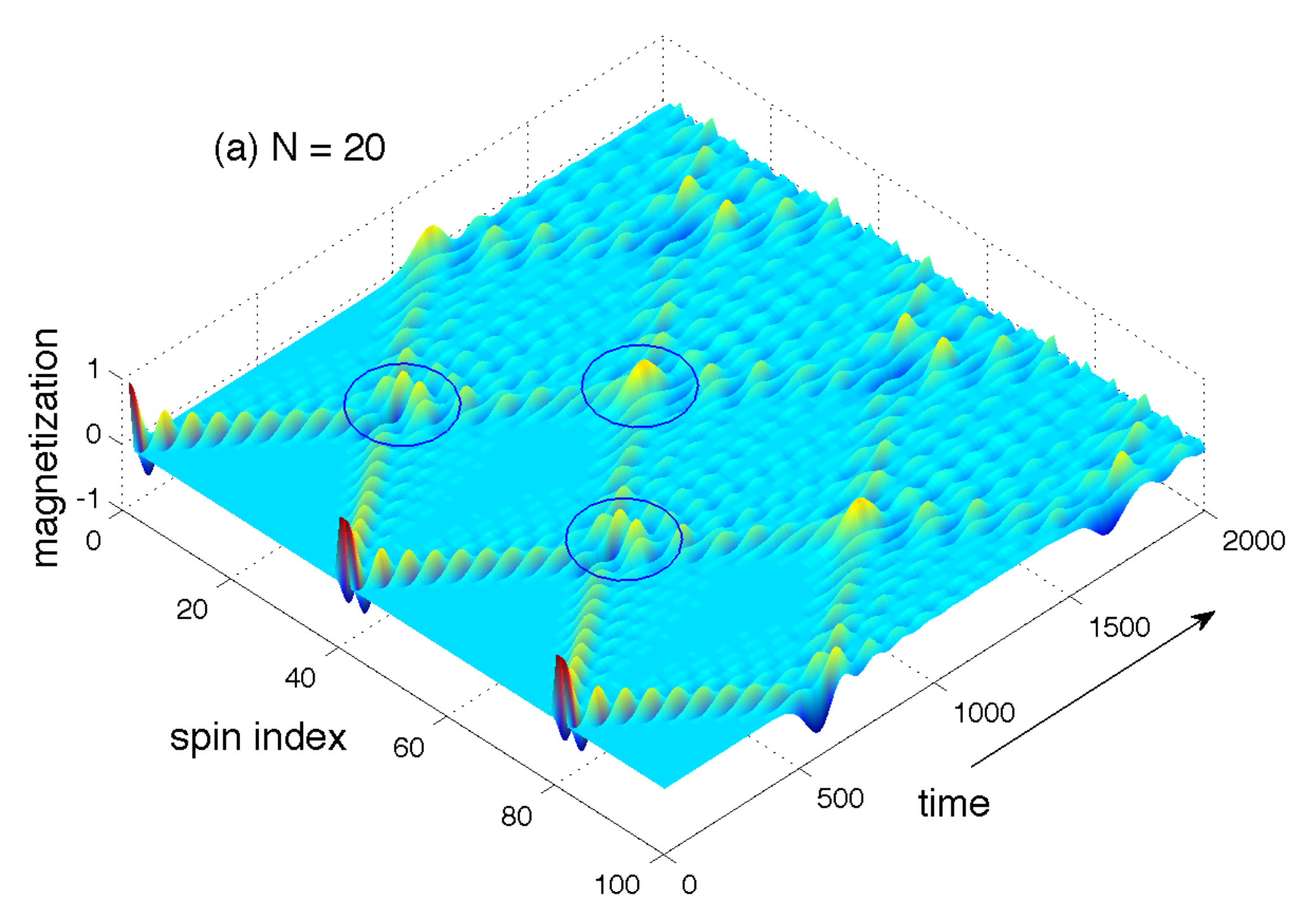}
	\includegraphics[scale=0.4]{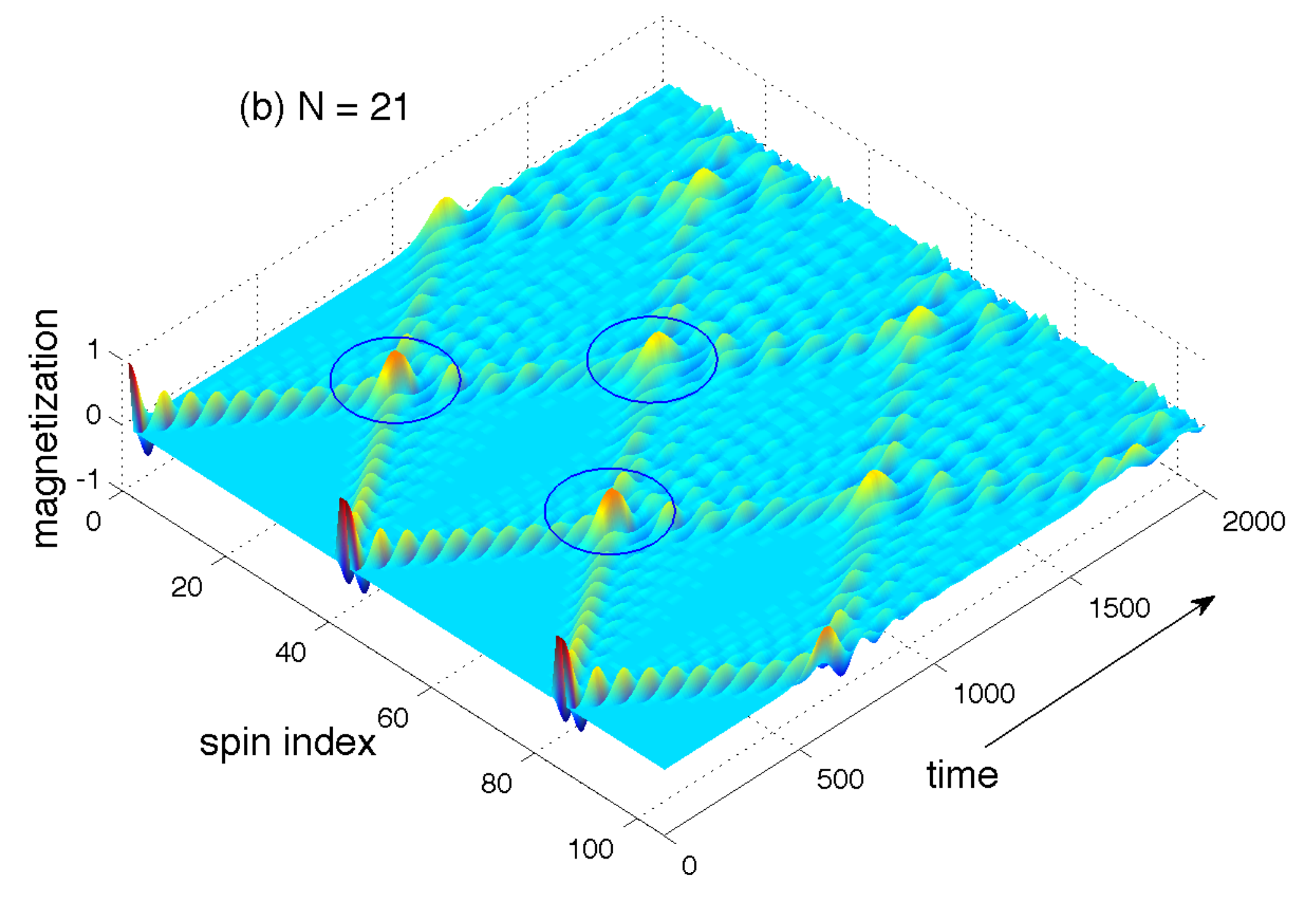}
\caption{Five copies of a spin chain illustrating the boundary
	interferences for (a) $N=20$; (b) $N=21$ during evolution
	under the NN DQ Hamiltonian.  The time axis is in units of
	$1/d$.}
\label{fig:DQinterferences}
\end{figure}

Fig.\ \ref{fig:DQinterferences} shows that the mirror enhancements
obtained under the DQ Hamiltonian are also due to interferences
generated as a polarization wave-packet reflects off the boundary in a
finite spin chain.  In the case of the DQ Hamiltonian the polarization
is inverted at every adjacent spin site as it travels down the chain,
and the nature of interferences at the boundary depends on whether the
chain contains an even or odd number of spins.

\begin{figure}[htb]
	\centering
		\includegraphics[scale=0.5]{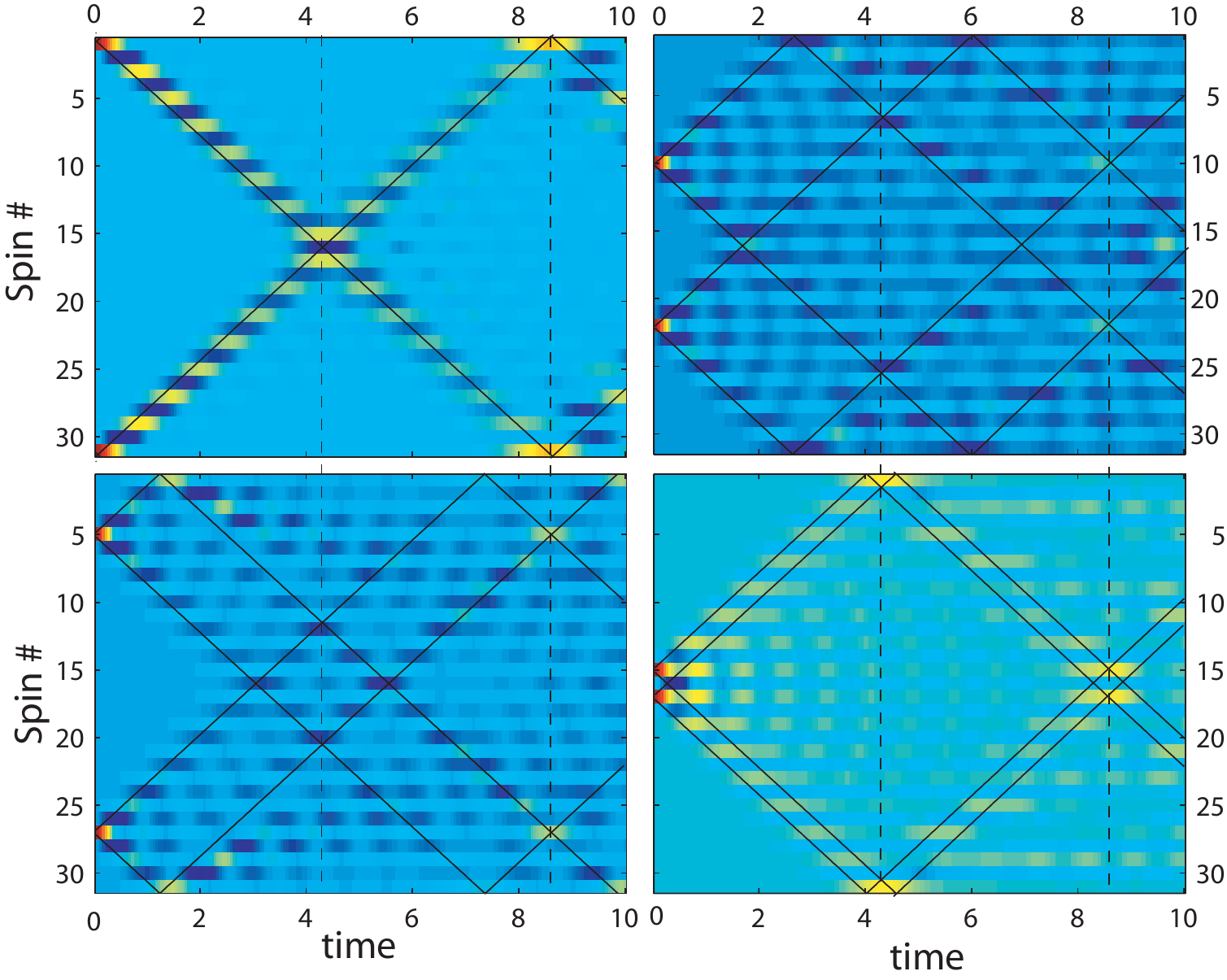}
	\caption{Zero quantum coherence intensities
	$J_{j,l}^0(t)+J_{N+1-jl}^0(t)$ as a function of time in a
	chain of $N=31$ spins. Left top: $j=1$; Left bottom: $j=5$;
	Right top: $j=10$; Right bottom: $j=15$. The black lines are
	drawn as a guide for the eye.}
	\label{JmqcPlot}
\end{figure}

The above mirror dependence on chain length is also manifested in the
multiple quantum coherence dynamics of the spin chain. Mirror times
were also observed in the multiple quantum dynamics for thermally
polarized spin chains \cite{Zhang-2009}.  We can explain this as
follows.  Although significant overlap for waves originating at
symmetric spin positions $j$ and $N+1-j$ can occur at different times
depending on $j$ (for example, at times $ t_{j} \propto N\pm j$) there
is also large overlap at the time $t_m\propto N$, when one of the
waves has bounced off a boundary and encounters the second wave
originating from the symmetric spin (see Fig. \ref{JmqcPlot}).  Since
at each time $t_j$ there occurs only the overlap for one pair,
weighted by a factor $N$, only the overlaps at the mirror time
$t_m\propto N$ are noticeable for the thermal initial state
\cite{Zhang-2009}.

\section{Summary}

In summary, we have measured the single-spin and two-spin
Liouville-space operators that drive the (nearly) ballistic transport
of magnetization under the DQ Hamiltonian at room temperature.  We
also developed a real-space description of the spin dynamics that is
able to accurately describe the observed results, by allowing us to
characterize the growth and coherent dynamics of multi-spin
correlations, and to explain the origin of the previously described
``mirror times'' in finite spin chains.  Within the validity of a NN
approximation, the dynamics of the $N$-spin system are seen to be
restricted to a Liouville space whose size grows only quadratically
with $N$.  Based on our model, we estimate that the magnetization is
coherently transported down the chain with a group velocity of
$6.04\pm0.38$ $\mu$m/s. We expect this velocity to be directly
independently measurable using reciprocal space NMR methods
\cite{Sodickson-1998}.


\ack{This work was funded in part by the NSA under ARO contract number
W911NF0510469 and the NSF under Awards 0702295, DMR-1005926 (to PC)
and PHY-0903727 (to LV). DGC acknowledges support from the Canadian
Excellence Research Chairs (CERC) program.  The FAp crystal used in
the experiments was grown by Prof. Ian Fisher.}


\appendix


\section{Evolution of the end-polarized state under $\mathcal{H}_{\mathrm{DQ}}$}

\begin{figure}[htb]
\centering
\includegraphics[scale=0.4]{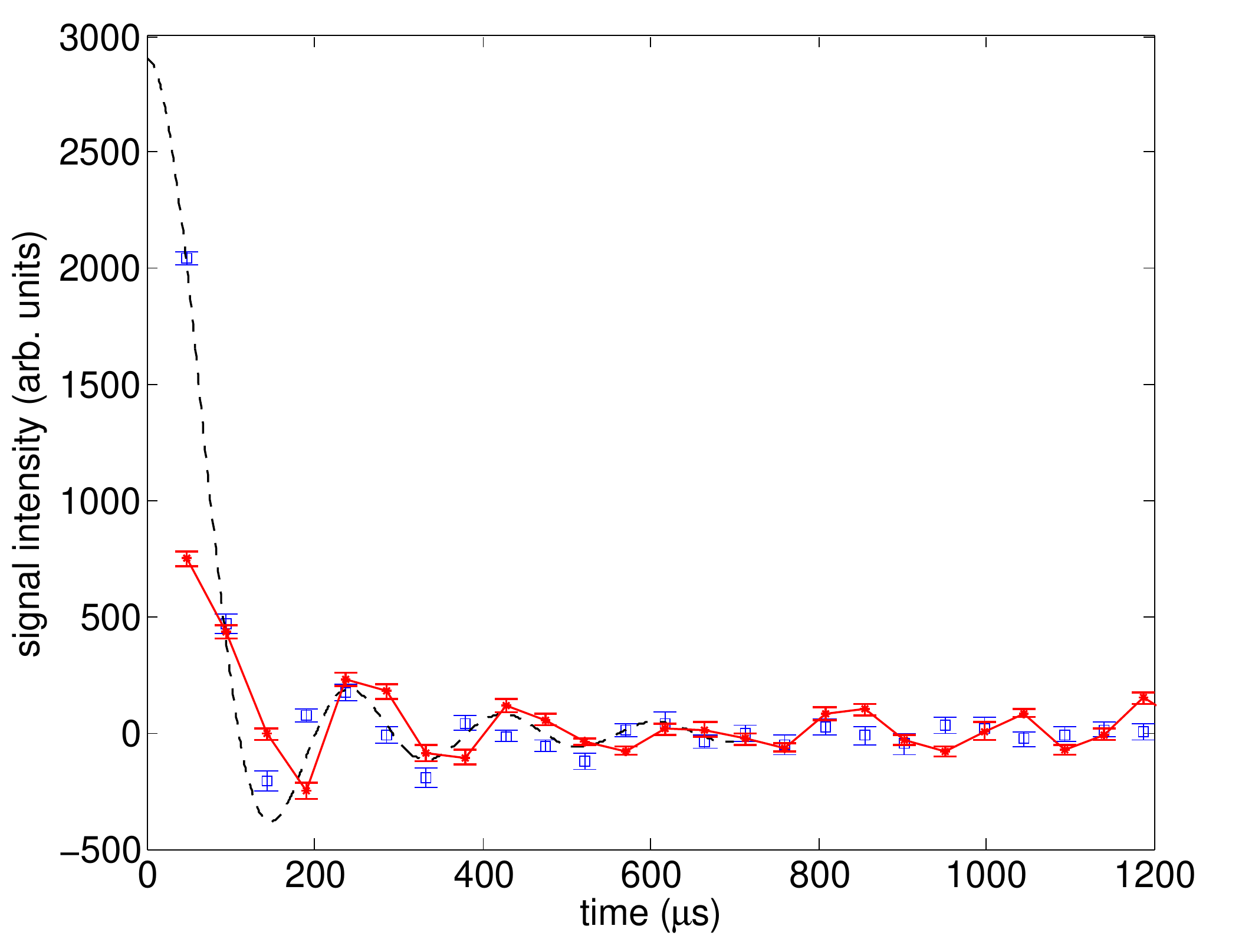}
\caption{Evolution of the end-polarized initial state
under $\mathcal{H}_{\mathrm{DQ}}$, showing the collective
magnetization ($\sum_i \sigma_z^i$ --- blue squares) and the two-spin
correlations ($\sum_{i} (\sigma_x^i\sigma_y^{i+1} +
\sigma_y^i\sigma_x^{i+1})$ --- red asterisks).  The dashed black line
is the best fit of the observed magnetization to the analytical model
described later (yielding $d = 8.7 \times 10^3$ rad/s). The solid red
line connecting the red asterisks is a guide to the eye.}
\label{fig:end-chain8}
\end{figure}

Figure \ref{fig:end-chain8} shows the experimental results obtained
using the 8-pulse implementation of the DQ Hamiltonian (in contrast to
the 4-pulse version shown in Figure 2), when the spins are prepared in
the end-polarized state $\rho_{1N} \approx \mathbf{1} -
\epsilon(\sigma_1^z + \sigma_N^z)$ in which the polarization is
localized at the ends of the chain.  Fitting the magnetization
dynamics to the analytical model results in a value of $d= 8.7 \times
10^3$ rad/s.

\section{Short time evolution}

Let the initial state of the spin system be $\rho_j = \mathbf{1} -
\sigma_z^j$.  The state of the spin system at a time $t$ later,
following evolution under the DQ Hamiltonian is
\begin{equation}
\hspace*{-0.5in}\rho_t = \rho_0 - it\left[\mathcal{H}_{\mathrm{DQ}},
\rho_0 \right] -\frac{t^2}{2}
\left[\mathcal{H}_{DQ},\left[\mathcal{H}_{\mathrm{DQ}},
\rho_0\right]\right] - \frac{it^3}{6}\left[\mathcal{H}_{\mathrm{DQ}},
\left[\mathcal{H}_{\mathrm{DQ}},
\left[\mathcal{H}_{\mathrm{DQ}},\rho_0\right]\right]\right] + \dots
\end{equation}
Evaluating the above commutators, we get
\begin{equation}
\left[\mathcal{H}_{\mathrm{DQ}},\rho_0\right] =
2d\left(h^{(2)}_{j-1,j} + h^{(2)}_{j,j+1}\right) \;,
\end{equation}
\noindent 
where $h^{(2)}_{i,j} = \sigma_i^+\sigma_j^+ - \sigma_i^-\sigma_j^- =
\frac{i}{2}\left(\sigma_i^x\sigma_j^y + \sigma_i^y\sigma_j^x\right) $.
A collective rotation of $h^{(2)}_{i,j}$ about the $z$-axis by an
angle $\phi$ results in phase shifts by angles $\pm 2\phi$, indicating
a double quantum coherence.  Similarly,
\begin{eqnarray}
\hspace*{-1in}\left[\mathcal{H}_{\mathrm{DQ}},\left[\mathcal{H}_{\mathrm{DQ}},\rho_0\right]\right]
& = & -2d^2\left(\sigma_{j-1}^z + 2\sigma_j^z+\sigma_{j+1}^z\right)
\nonumber \\
& - & 2d^2
\left(h^{(3)}_{j-2,j-1,j} + 2h^{(3)}_{j-1,j,j+1} +
h^{(3)}_{j,j+1,j+2}\right),
\end{eqnarray}
where $ h^{(3)}_{i,j,k} = \sigma_i^+\sigma_j^z\sigma_k^- +
\sigma_i^-\sigma_j^z\sigma_k^+ $ and
\begin{eqnarray}
\hspace*{-1in}\left[\mathcal{H}_{\mathrm{DQ}},\left[\mathcal{H}_{\mathrm{DQ}},\left[\mathcal{H}_{\mathrm{DQ}},\rho_0\right]\right]\right]
 & = & 6d^3\left(h^{(2)}_{j-2,j-1} + 3h^{(2)}_{j-1,j} +
 3h^{(2)}_{j,j+1} + h^{(2)}_{j+1,j+2}\right) - \\ & &
 \hspace*{-0.5in}2d^3\left(h^{(4)}_{j-3,j-2,j-1,j} +
 3h^{(4)}_{j-2,j-1,j,j+1} + 3h^{(4)}_{j-1,j,j+1,j+2} +
 h^{(4)}_{j,j+1,j+2,j+3}\right), \nonumber
\end{eqnarray}
where $ h^{(4)}_{i,j,k,l} = \sigma_i^+\sigma_j^z\sigma_k^z\sigma_l^+ +
\sigma_i^-\sigma_j^z\sigma_k^z\sigma_l^- $, and so on.  Collective
rotations of $h^{(3)}_{i,j,k}$ and $h^{(4)}_{i,j,k,l}$ about the
$z$-axis by an angle $\phi$ results in phase shifts by angles $0$ and
$\pm 2\phi$ respectively, indicating these are zero and double quantum
coherences.  We observe the same alternating zero and double-quantum
signatures in the commutator expansion as obtained in the fermionic
solution (Eq. (3) in the main text). Moreover, here we see that the
growth of the spin system is necessarily slow.  At the level of the
second commutator, we can infer that the system is 3 times more likely
to evolve from a 3-spin state to a 2-spin state, than it is to evolve
to a 4-spin state. This is one indication of why locally polarized
states are transported almost ballistically along the spin chain.
This is schematically illustrated in Fig.\ B1.

\begin{figure}[thb]
\centering
\includegraphics[scale=0.4]{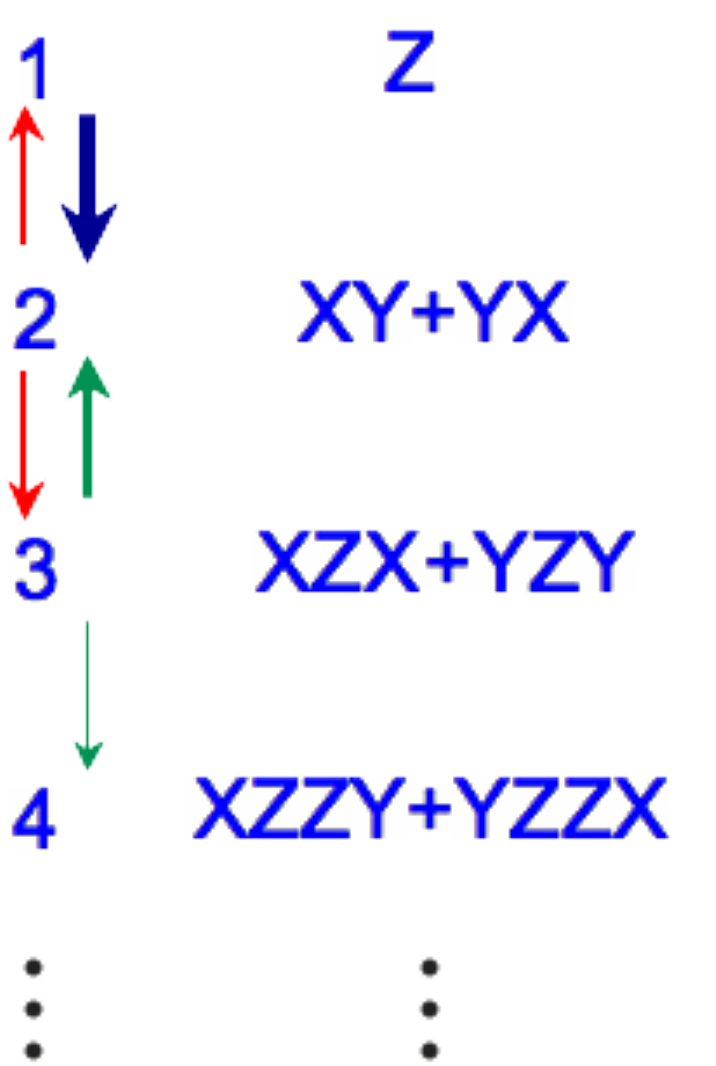}
\caption{The successive commutators in Eqs.\ B1--B4 yield the
Liouville space operators shown in the figure and illustrate the
progressive generation of zero and double quantum coherences.}
\label{fig:model1}
\end{figure}


\section*{References}


\bibliography{Bibliography}




\end{document}